\documentclass[
aps,%
10pt,%
final,%
notitlepage,%
oneside,
 onecolumn,%
nobibnotes,%
nofootinbib,%
superscriptaddress,%
noshowpacs,%
centertags]%
{revtex4}

\begin{document}
\selectlanguage{english}

\title{Angular Size-Redshift: Experiment and Calculation}

\author{\firstname{V.R.}~\surname{ Amirkhanyan}}

\affiliation{ Sternberg Astronomical Institute, Lomonosov Moscow
State University, Moscow, 119992 Russia}
\begin{abstract}
In this paper the next attempt is made to clarify the nature of
the Euclidean behavior of the boundary in the angular
size-redshift cosmological test. It is shown experimentally that
this can be explained by the selection determined by anisotropic
morphology and anisotropic radiation of extended radio sources. A
catalogue of extended radio sources with minimal flux densities of
about 0.01 Jy at 1.4 GHz was compiled for conducting the test.
Without the assumption of their size evolution, the agreement
between the experiment and calculation was obtained both in the
$\Lambda CDM$ model ($ \Omega_m=0.27$,$\Omega_v=0.73$.) and the
Friedman model ($\Omega = 0.1 $).\\
 DOI: 10.1134/S1990341314040026
\keywords{Keywords: cosmology: observations-cosmological
parameters}
\end{abstract}
\maketitle
\section{INTRODUCTION}

Apparently, Hoyle [1] was the first to note that in the Friedman
space with q > 0, the angular size $\Theta$ of an object with the
extent of D at a redshift equal to about 1 attains a minimum.
Thereby, the possibility to estimate the space geometry emerges.
The first to conduct this test was Legg [2], who showed the
relation between the 3CR catalog extended radio source angular
sizes and the redshift. It turned out that the upper limit of the
diagram, which corresponds to the longest extent of the radio
sources of 400 kpc, shows the best agreement with the Euclidean
geometry, not with the Friedman one. Realizing that this is
absolutely impossible, Legg assumed that the linear sizes of the
radio sources decrease with increasing redshift. In the
Einstein-de Sitter model (q = 0.5), he obtained the best agreement
with the observations assuming the size evolution $D \sim
(1+z)^{-1.5}$. In the same work Legg wrote prophetically that we
observe not the true size of the radio source but its projection
on the celestial sphere. Thus he asserted that the structure of
the radio source is anisotropic and randomly oriented. Miley [3]
and Hooley et al. [4], who used a wider sample of extended radio
sources, obtained the same result. The most representative sample
of 540 radio sources was used by Nilsson et al. [5] for the
plotting the $\Theta$-z diagram. It is mostly composed of the
radio sources of the 3CR and 4C surveys. This work confirmed the
predecessors' results: the boundary of the $\Theta$-z diagram is
proportional to 1/z. In 1982 Orr and Brown [6] assumed that the
radio sources radiation pattern is not spherical: its maxima
coincide with the direction of the jet. The authors constructed
the diagram as axially symmetric to the radio source major axis
which allowed them to make it a function of one argument: the
orientation angle of the symmetry axis relative to the observer.
This fruitful idea allowed them to abandon two types of radio
sources (compact and extended) and consider the statistics of
randomly oriented extended objects. At the conference in 1987
Amirkhanyan [7] showed that the non-spherical morphology, and the
non-spherical radio source radiation connected with it, is
inevitably accompanied by a simple selection effect, which imposes
an upper limit on the apparent angular size of distant radio
sources.
\section{SELECTION}
If the radiation and the structure of the object are not
spherical, then the observed flux density S and the angular size
$\Theta$ are defined not only by the luminosity Lv, the physical
dimension $D_0$, and remoteness of an object $l_v$ but also by its
orientation relative to the observer. The telescope can "see" the
object if its flux density S exceeds the detection threshold
$S_t$:
\begin{equation}
    S=\frac {L_v\varphi(\phi)} {l_v^2} > S_t ,
    \end{equation}

here $\varphi(\phi)$ describes the form of the radiation pattern.
When $\phi=0$, the radiation toward the observer is
maximal, and when $\phi=\pi/2$, it is minimal. It follows
from the detection condition (1) that such a

radio source will be seen whatever the orientation
($\phi=0\div\pi/2$  ), up to the distance
\begin{equation}
l_t=\sqrt{\frac {L_v\varphi(\pi/2)} {S_t}}
\end{equation}
The selection appears with the distance increase,
because the orientation angle of the radio source
at which it could be observed decreases. Consequently,
its projection on the celestial sphere decreases
($D=D_0 \sin\phi $). The upper boundary of the
$\Theta$ - z diagram, which is usually calculated on the tacit
assumption of spherically symmetric radio source
radiation, at the distances greater than $l_t$ gets one
more restriction in the form of the multiplier $\sin\phi_{max} $.
Knowing the form of the radio source radiation
pattern [7, 8], one can easily obtain the dependence of
the maximum angular size on redshift in the assumed
space model and with a defined detection threshold.
Such calculations were carried out in [7] for the
3CR catalogue, where the detection threshold is equal
to 10 Jy at the frequency of $\nu $ = 178 MHz. As a
result, the upper limit of the $\Theta$ - z diagram close to
1/z was obtained in the Einstein.de Sitter model. In
order to explain the selection mechanism, let us show
the dependence of the orientation angle boundary on
the redshift (see Fig. 1), plotting the radio source
radiation pattern from [8]
\begin{equation}
 \varphi(\phi)= \frac {L} {L_0} = a(1+z)^{\alpha} + (1-a)
\cos^{2n}\phi .
\end{equation}
Here $L_0$ is the luminosity toward the radiation peak; $a$  is
the ratio of the luminosity of the isotropic component of the
radiation to $L_0$ in the comoving coordinate system; $\alpha$ is
the spectral index of the isotropic component; $\phi$ is the angle
between the orientation of the radiation maximum and the
observer's line of sight; n specifies the width of the main lobe
of the radiation pattern. The calculations were done for the
Friedman model with q = 0.05. The spectral index is
$\alpha=-0.85\, (S \sim \nu^{\alpha}) $. We applied the parameters
of the radiation pattern derived in this paper: a = 0.005, n = 15.
The black line corresponds to the minimum flux density of 2 Jy of
the catalogue, the blue line is for 0.05 Jy. In Fig.1 there is a
flat region corresponding to arbitrary radio source orientations
from 0. to 90.. The length of this region naturally increases with
the decrease of the detection threshold. The extension of the
redshift beyond the flat region is accompanied by the decrease of
the maximum allowed orientation angle $ \phi_{max} $ and, as a
consequence, by the radio source selection by this angle: the
observer detects radio sources with the orientation less than $
\phi_{max} $, and this results in the limitation of the observed
angular size. In view of the above, the routine question arises -
what actually determines the apparent boundary of the $\Theta - z$
diagram: the cosmological evolution of the radio source sizes or
the selection by the orientation angle?
\begin{figure*}[tbp]
\setcaptionmargin{5mm} \onelinecaptionsfalse
\includegraphics[scale=.7]{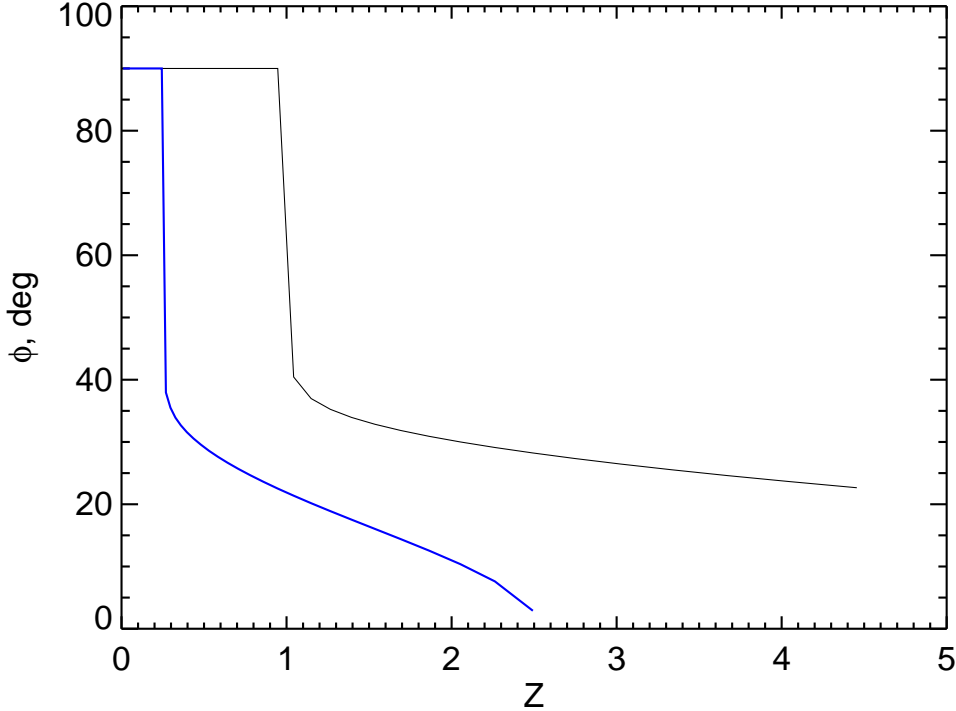}
\caption{ The dependence of the orientation angle's boundary from
redshift for the two detection thresholds. Blue line- $S_t $= 2
Jy, black line - $S_t $= 0.05 Jy } \label{fig_fi_z}
\end{figure*}

 \section{CATALOG}
To solve this dilemma, it is enough to reduce the detection
threshold $S_t$, and thereby to defer the hypothetical selection
to a higher z. If the radio sources stay under the boundary 1/z,
it definitely means evolution. Hence the simple and natural
conclusion follows: an extensive catalog of weak extended radio
sources is needed. In order to defer the selection to $z \approx
1$, $S_t$ should be reduced by more than ten times. As there was
no such catalog, it had to be made. It was compiled using the
following data. \\1. The catalog of double radio sources from
Nilsson et al. [5]. Several extended 3CR radio sources were added
to it. Note that in [5] the boundary is calculated for the radio
sources' maximum size of 4 Mpc assuming that the Hubble constant
is $H_0$ = 50 km/s/Mpc. For the modern $H_0$ = 71 km/s/Mpc/ it
corresponds to 2.8 Mpc. Further this value is used to calculate
the $\Theta$ - z diagram boundaries.\\
 2. Radio sources with
known z from Machalski et al. [9]. An extended radio source from
Machalski et al. [10] was added to this list.\\ 3. Double QSOs
from Buchalter et al. [11].\\ 4. Extended radio galaxies at
declinations greater than 60. from Lara et al. [12, 13].\\ 5.
Double SUMSS radio sources from Saripalli et al. [14].\\ 6.
Double QSOs from Amirkhanyan [15].\\

Moreover, by means of the modified programs from [15], the NVSS
catalog [16] was analyzed, more than two thousand candidates for
extended radio sources were formed and identified with the NED and
SDSS optical objects. The programs are set up for searching the
objects grouped by the word "double" even if their morphology is
more complex. These programs cannot always cope with their task
successfully if the object extent exceeds $200''$, as the
probability of the projection of false radio and optical
components onto the radio source image increases. Aiming to
minimize the number of possible errors, the author looked through
NVSS and FIRST [17] maps of the radio sources in which optical
components with measured redshifts were found. If their morphology
does not meet our concept of a classical "double" structure, they
are excluded from our catalog. The redshifts of nine formed radio
sources have been measured at the 6-m telescope of the Special
Astrophysical Observatory (the paper is being prepared).
Sixty-seven objects from this list crossed the conditional limit
of 1 Mpc (in the $\Lambda CDM$ space model currently accepted by
the astronomical community) and noticeably expanded the giants
list. As a result, 599 objects were added to the above mentioned
lists. The total number of radio sources in the joint catalog is
1953: radio galaxies-913, quasars-1040. The minimum flux densities
are about 0.01 Jy at the frequency of 1.4 GHz. If the radio source
occurs in several of the above mentioned catalogs, the first
publication is given the priority. The parameters (mainly redshift
and angular size) of some of them are corrected according to the
modern data. Some examples. The angular size of the 3C270 radio
source in the Nilsson et al. [5] catalog is $498''$, and on the
NVSS map its size is no less than $3000''$ . The 3C449 object has
the size of no less than $3600''$ on the NVSS map, although its
catalog size is $300''$. The 3C263.1 redshift is 0.824 [18], not
0.366.
\section{ OBSERVATIONS AND CALCULATIONS}
Let us take the $\Theta - z$ diagram from [5] (Fig. 2) and

\begin{figure*}[tbp]
\setcaptionmargin{5mm} \onelinecaptionsfalse
\centerline{\includegraphics[scale=0.7]{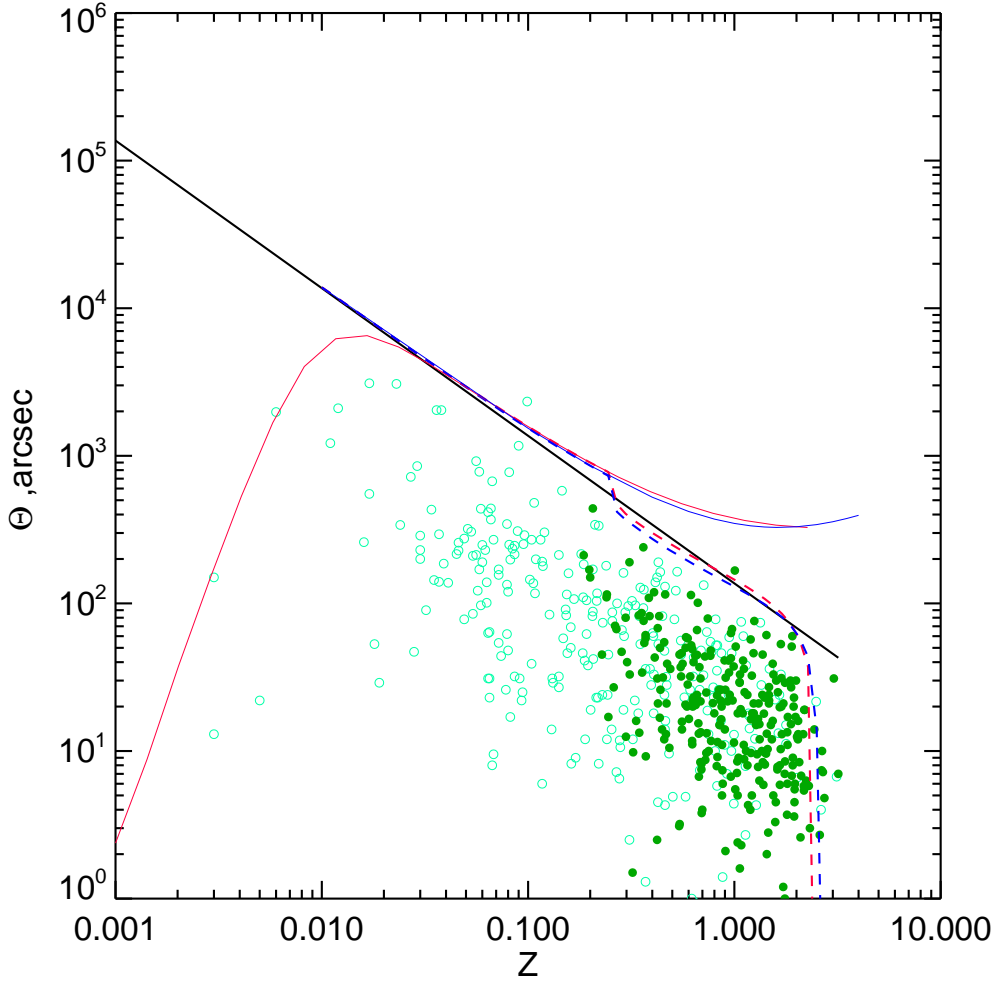}}
\caption{Angular size-redshift of Nilsson et al. [5]. Key: open
cyan circles-radio galaxies, filled green circles - QSOs. Black
straight line - boundary of the Euclid static model. Red curve
-standard boundary of the Friedman model, $\Omega= 0.1$ . Red
dashed curve - boundary of the Friedman model with selection. Blue
curve - standard boundary of the $\Lambda CDM$ model ($\Omega
_m=0.27$ ,$\Omega _v=0.73$). Blue dashed curve - boundary of
the $\Lambda CDM$ model with selection.\\
 }
 \label{fig_t_z1}
\end{figure*}

\begin{figure*}[tbp]
\setcaptionmargin{5mm} \onelinecaptionsfalse
\centerline{\includegraphics[scale=.7]{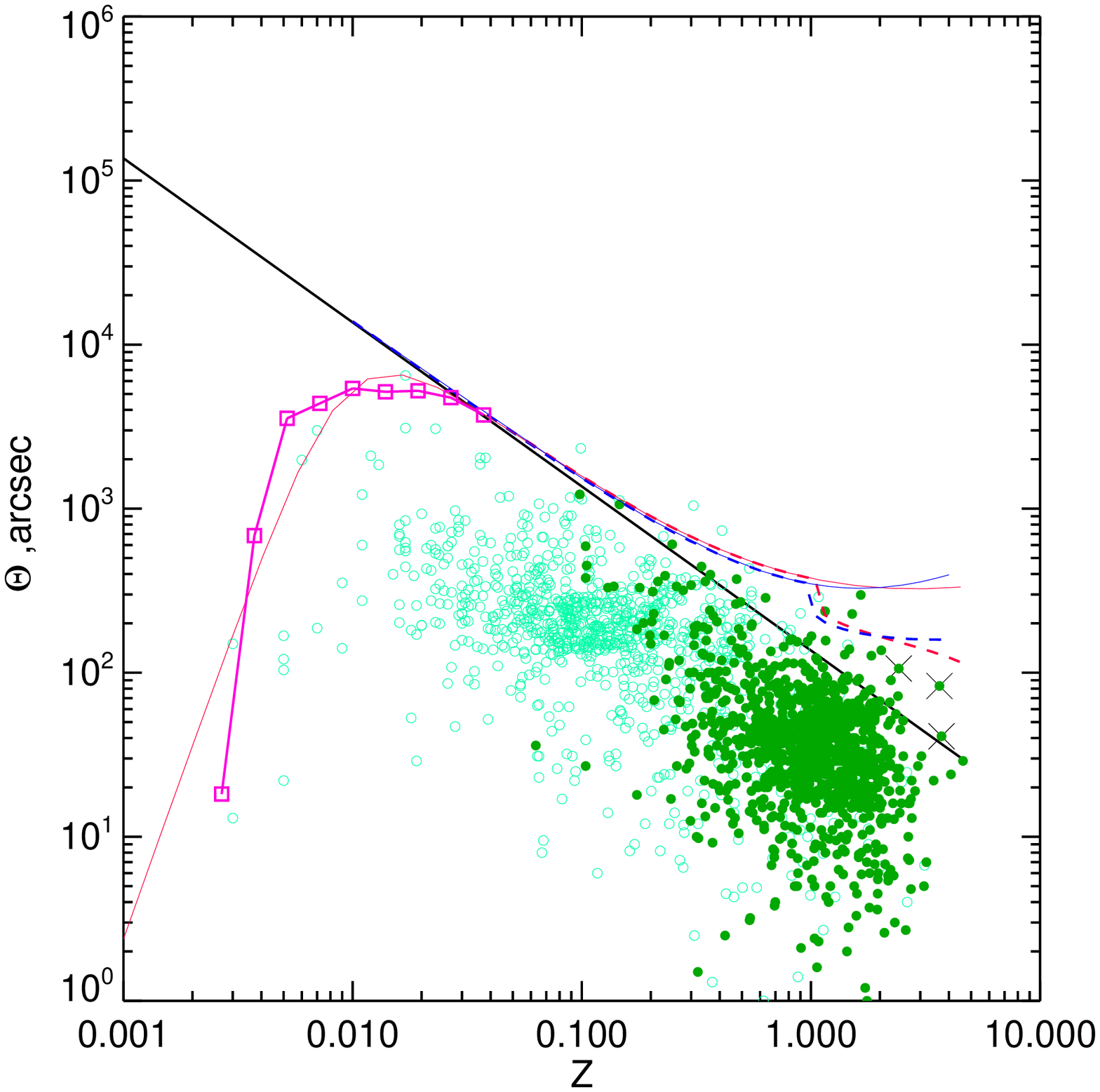}} \caption{
Angular size-redshift of the full catalog. Magenta
squares+line-calculated boundary at z < 0.1.
Other subscripts are as in Fig. 2.\\
 }
\label{fig_t_z2}
\end{figure*}

calculate its upper limit allowing for selection. For
that we take the average model of the radio source
radiation pattern from [8]. The integral of the diagram
(3) over the full sphere differs from the standard
$4\pi$ [19]:
\begin{equation}
      h=4\pi a(1+Z)^{\alpha}+2\pi (1-a)\frac {(2n-1)!!} {(2n)!!}
\end{equation}

Hence, the full radio source luminosity
          $$L_s=hL_0  $$
The author experimented with several models of the radiation
pattern, but this did not add any fundamental changes to the final
result. The following parameter values have been used while
calculating: the Hubble constant $H_0$ = 71 km/s/Mpc; the largest
radio source size $D_{max}$ = 2.8 Mpc; the highest radio source
luminosity $L_{max} = 2 \times 10^{28}$ W/Hz (this value
corresponds to the highest radio source luminosity of the present
catalog); deceleration parameter in the Friedman model q = 0.05
($\Omega= 0.1$). In the present $\Lambda CDM$ model
 $ \Omega_m=0.27$,$\Omega_v=0.73$.\\ In addition to the radio
sources, Fig. 2 shows: \\
1. relation $\Theta$ - z in the Euclid static model (black
straight line)\\
2. $\Theta$ - z relation in the Friedman model without the
selection (red curve)\\ 3. $\Theta$ - z relation in the Friedman
model with the selection (red dashed curve)\\ 4. $\Theta$ - z
relation in the $\Lambda CDM$ model without the selection (blue
curve)\\ 5. $\Theta$ - z relation in the $\Lambda CDM$ model with
the selection (blue dashed curve).\\
The calculated selection adequately agrees with the upper limit of
the radio source angular sizes in a wide range of radiation
pattern parameters (3): $a$=0.002 $\div$ 0.01, n = 10 $\div$ 20.
At that, the width of the radiation pattern at the 0.5 level
ranges within 30 $\div$ 21. The best result can be achieved at $a$
= 0.005 and n = 15 (Fig. 2). The detection threshold $S_t =2.0 $
Jy corresponds to the depth of the 4C survey [20, 21]. It is
clearly seen that the radio sources, as shown in [5], really lie
below the "Euclid boundary"
  $$  \Theta_E(z)= \frac {D_{max} } {\frac {cz}{H_0}} .$$

The selection, which starts at z > 0.25, pressed the boundaries
calculated by the standard formulas (lines 2 and 4) to the
"Euclidean boundary" which agrees with the observations. Figure 3
shows the $ \Theta -z $ diagram of the fully compiled catalog
(subscripts identical to Fig. 2). Unlike Fig. 2, this diagram
shows that the angular sizes of 43 radio sources overcame the
"Euclid boundary" and are limited by the function $ \Theta -z $
calculated in terms of selection (lines 3 and 5). The detection
threshold here is determined by the deepest surveys.FIRST and
NVSS. The minimum flux densities of the radio sources based on
these surveys are $ 0.007 \div 0.015 $ Jy which is in accordance
with the threshold $ S_t \approx 0.05 $ Jy at $\nu$ = 178 Mhz. The
other parameters are the same as for Fig. 2. Hence, with the
detection threshold decreased from 2 Jy to 0.05 Jy, the selection
shifted from z = 0.25 to z = 0.95 (Fig. 1). This allowed the
observer to discover weak extended radio sources, the angular
sizes of which at z > 0.05 agree sufficiently with the $\Theta -z
$ dependence in the standard space models. Naturally, such a
result casts doubts on the obtained scale of the cosmological
evolution of the radio sources size. The list of the radio sources
which exceeded $\Theta_E(z) $ is given in the table 1. The
doubtable objects have question marks. These three objects
sufficiently agree with our concept of classical "double" radio
sources. At the same time the author cannot rule out that they can
be formed by the odd projection of the physically unrelated
optical and radio components. In the redshift region less than
0.05, where the radio galaxies are basically situated, both in
Fig. 2 and in Fig. 3 the radio source maximal angular sizes are
orders of magnitude smaller than the calculated ones. This is
easily explained by the decrease of the space volume, and,
consequently, by the decrease of the probability of the detection
of an extended galaxy. This can be seen from the equation:
\begin{equation}
    N=\rho\Delta V(z) \int\limits_{D_z}^{D_{max}} p(D)dD
\end{equation}

Here $p(D)\,dD$ is the probability density function of
the radio galaxy linear sizes, N is the number of
galaxies larger than $D_z$ in the volume $ \Delta V(z)$ at their
space density $ \rho $ .\\
It is important that despite the 40 times reduction
of the catalog minimum flux densities and the
growth of the number of radio galaxies, the angular
size boundary in this redshift region did not change.
Hence, radio astronomers have catalogued the majority
of close and most extended radio galaxies, and
the angular size limitations at z < 0.05 are close to
realistic. Let us use this boundary for the evaluation of


the space density of the radio galaxies. The upper limit of their
linear sizes at some z and consequently their maximum angular size
as a z function, are defined by solving the equation (5) for $ D_z
$. Let us assume the upper limit criterion as the condition that
no more than one object can exceed the $D_z$ level, and define N =
1. In such a case it is not necessary to assume a more strict
condition, defining N <1. To solve the equation, we need to know
the probability density distribution function of the radio galaxy
linear sizes in the z < 0.1 region. We will define it in two ways:
simple and not so simple.
 Let us start with the following
assumptions: the space density of the radio galaxies and their
linear size distribution function do not depend on the redshift.
There are 392 radio galaxies with z < 0.1. The linear size
distribution function of this sample, calculated with the help of
the standard procedures, is shown in Fig. 4 (black curve). This
very function we put under the integral in the equation (5).
Further we assign the space density, divide the z range into
several intervals, and, operating with the lower limit of the
integral, solve the equation (5) in each interval. As a result, we
obtain the dependence of $ D_z $ on z. We change the space density
$ \rho$, then repeat the calculations and achieve the best visual
fit of this curve with the radio galaxies apparent boundary in
Fig. 3 (squares connected with the bold line). The best result is
achieved for $ \rho =10^{-4} Mpc^{-3}$. The second way is simple.
If we put the distribution function in the form
$$p(D)dD \sim D^{\gamma} dD $$,
then the equation (5) can be easily solved relative to $ D_z $. We
get a simple but cumbersome formula with which the $\Theta - z$
boundary was calculated, Figs. 2 and 3 (red curve). In this case,
in order to adjust the calculations to the experiment we had to
manipulate the parameters $\Theta - z$ and $\gamma $. In the
author's subjective opinion the best result was achieved for $
\rho =10^{-4} Mpc^{-3}$ and $ \gamma  =-1.6 $. If we adjust this
boundary to the first variant, we need to define the parameters $
\rho =8\cdot 10^{-5} Mpc^{-3}$ and $ \gamma  =-1.1 $. The derived
density estimate relates to the radio galaxies the luminosity of
which is within the $3\cdot 10^{21} \div 1.5\cdot 10^{25}$ W/Hz
range.\\
 Let us focus on Fig. 4 again. The shown experimental
probability density distribution function of the radio galaxy
linear sizes is the distribution function of the {\em visible}
sizes D. It is the result of the convolution of two functions: the
real sizes distribution function $D_o$ and the distribution
function of the radio galaxies orientation angles $\phi$ relative
to the observer. The distribution function of these angles at a
random space orientation of the radio sources is $ \sin\phi\,d\phi
$. Let the real sizes of the objects be between $D_{min}$ and
$D_{max}$. Then, taking into account that $ \sin\phi=\frac{D}{D_o}
$ , we obtain
\begin{equation}
    p(D)dD= DdD \int\limits_{D}^{D_{max}} \frac{P(D_o)dD_o}{D_o
    \sqrt{({D_o}^2-{D}^2)}} .
\end{equation}

If $D<D_{min}$ then $ p(D)dD \approx DdD $,  and depends on the
distribution of the real sizes only weakly. Hence, the maximum of
the distribution of the visible sizes is near $D_{min}$. As the
distribution function of the visible sizes in Fig. 4 has its
maximum in the region of 0.2 $\div$ 0.4 Mpc, one would expect that
the minimum true sizes of the radio sources from this sample are
in the same range. The catalog has been generated of 10000
objects, whose sizes are randomly distributed according to the $
D_o^{-2.5} dD_o $ law within 0.25 $\div$ 3.5 Mpc; their
orientation angles comply with the distribution $ \sin\phi\,d\phi $ \\
Let us multiply the size of each object by its orientation angle
sine, and we get the catalog of their apparent sizes. The
distribution function of these sizes is shown in Fig. 4 (blue
line). Its form in the range of D >Dmin is close to the real sizes
distribution function. A reasonably good agreement between the
distributions of the real and generated catalogs allows us to
conclude cautiously that the minimum true size of the radio
galaxies from the used list is close to 0.25 Mpc.
\begin{figure*}[tbp]
\setcaptionmargin{5mm} \onelinecaptionsfalse
\centerline{\includegraphics[scale=.7]{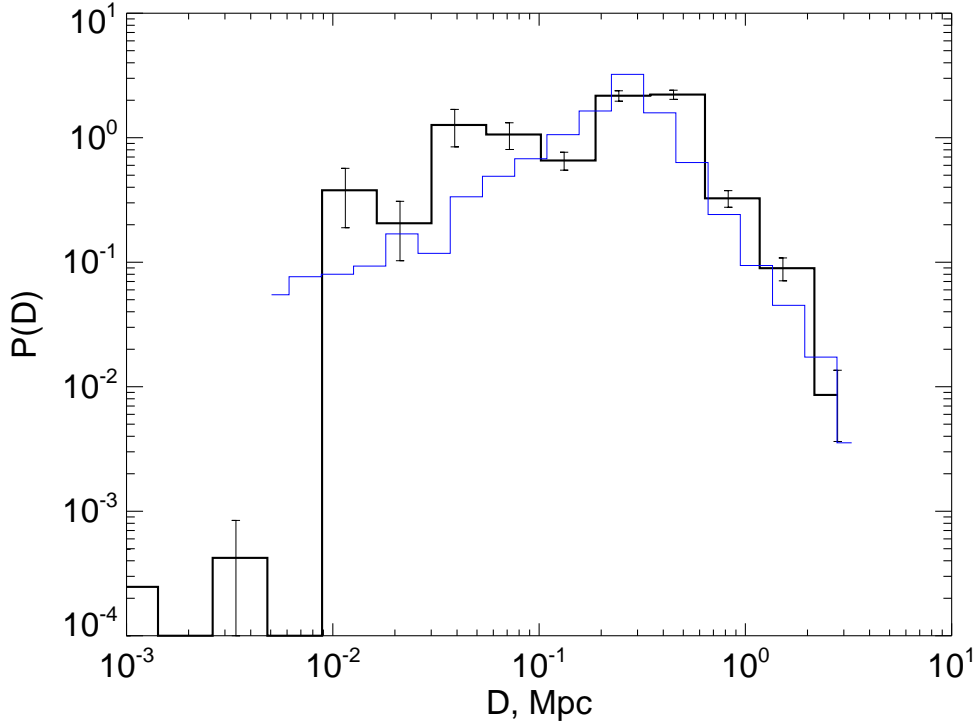}} \caption{
Distribution function of the visible sizes of the radio galaxies
(black line) and of the generated objects  (blue line).
 } \label{fig_pd}
\end{figure*}\\
\bigskip
\vspace{3cm}
\\
 table 1

   $\alpha_{2000}$ \hspace{0.8cm}  $\delta_{2000}$ \hspace{1.1cm}
   $\Theta ''$ \hspace{0.5cm} $ z $ \hspace{0.4cm} $ D (Mpc) $
   {\tt
{\footnotesize
\begin{verbatim}
Nilsson at al. [5]
  10 06 01.74  +34 54 10.4   2332  0.0990   4.21
  14 32 15.54  +15 48 22.4    167  1.0050   1.34
 Lara  at al.[12,13]
  07 50 34.80  +65 41 26.6    192  0.7470   1.41
  08 26 01.00  +69 20 37.0    432  0.5380   2.73
  19 51 40.80  +70 37 40.0    312  0.5500   1.99
  21 45 30.90  +81 54 53.7   1122  0.1460   2.84
FIRST Amirkhanyan [15]
  08 38 13.32  +13 58 07.3    137  2.0280   1.16
  08 56 25.56  +10 20 33.8     43  3.7150   0.30 ?
  10 00 13.03  +10 21 52.4     83  3.6390   0.61
  10 20 03.53  -02 47 22.9     95  1.4470   0.81
  11 07 09.83  +05 47 43.6     78  1.8000   0.67
  11 19 27.99  +13 02 49.3     72  2.3980   0.59
  12 14 31.40  +18 28 13.8     89  1.5900   0.76
  13 23 31.02  +54 59 50.3     66  2.2080   0.55
  13 33 06.10  +04 51 05.1    129  1.4050   1.10
  13 55 59.86  +19 04 14.4     90  2.2340   0.75
  14 03 26.30  +25 26 36.0     13  2.7560   0.47
  14 37 48.27  +07 48 38.0    139  1.4720   1.19
  14 39 33.69  +45 50 15.0    127  1.8360   1.08
  14 50 39.51  +45 49 50.7     92  1.6220   0.79
NVSS Amirkhanyan
  03 11 54.60  -31 30 14.1    106  2.4170   0.87 ?
  04 22 18.53  +15 12 39.3    735  0.4090   3.98
  07 53 39.25  +34 30 30.2    252  0.5480   1.61
  09 12 37.70  +68 33 55.2    290  1.0800   2.37
  09 39 38.16  -25 15 44.1    230  0.9000   1.80
  11 04 47.73  +21 03 13.2    237  1.1530   1.96
  11 30 20.00  -13 20 52.3    286  0.6340   1.96
  14 41 24.20  -34 56 41.5    157  1.1590   1.30
  15 13 39.91  -10 10 39.1    228  1.5130   1.95
  15 29 17.51  +32 48 35.8    298  1.6500   2.55 ?
  16 03 34.38  +36 59 43.0    155  0.9670   1.24
  21 56 41.59  -05 57 31.9    148  1.4450   1.26
Machalski at al. [9,10]
  07 25 17.40  +30 25 36.0    175  0.7900   1.31
  12 00 50.50  +34 49 21.0    147  0.5400   1.74
  12 54 34.00  +29 33 41.0    295  0.5500   1.88
  13 42 54.50  +37 58 18.0    678  0.2270   2.45
  14 45 25.20  +30 50 55.0    344  0.4170   1.87
  16 04 19.70  +37 31 17.3    182  0.8140   1.38
Saripalli et al.[14]
  02 37 09.90  -64 30 02.2    396  0.3640   1.99
  03 31 39.80  -77 13 19.3   1062  0.1460   2.69
  13 35 59.70  -80 18 05.1    606  0.2480   2.34
  17 28 28.11  -72 37 34.9    372  0.4740   2.20
  19 46 50.50  -82 22 53.8    444  0.3330   2.10
\end{verbatim}}}

\section{CONCLUSIONS}
New observed data proves that the anisotropy of the radiation and
structure of the radio sources cause a selection effect, placing
an upper limit on the apparent angular sizes. If a depth of a
survey is insufficient, the selection presses the boundary of the
angular size - redshift test to the "Euclidean boundary". As a
result, we see a contradiction between the boundaries in the
standard models and in the observed data. Decreasing the catalog
minimum flux density from 2.0 Jy to about 0.05 Jy allowed us to
shift the boundary of the selection by a factor of 4 $\div$ 5 in
the direction of increasing z. As a result, without the assumption
of the cosmological evolution of the extended radio source
physical sizes, an agreement was obtained between the observed
$\Theta - z $ test boundary and the calculated one both in the
Friedman model with q = 0.05 ( $ \Omega =0.1 $ ) and in the
$\Lambda CDM $ model
with $ \Omega_m=0.27 $ , $ \Omega_v=0.73$ .\\
The analysis of the angular size boundary at z < 0.1 allowed us to
evaluate the space density of close radio galaxies $ \rho \approx
10^{-4} Mpc^{-3}$.
\begin{acknowledgments}
The author would like to thank D. V.Amirkhanyan for the
substantial help in the work with the radio source images.
\end{acknowledgments}

Translated by N. Oborina

\end{document}